\documentclass[]{article}
\usepackage{color}
\usepackage{graphicx}    % Add some packages for figures. Read epslatex.pdf on ctan.tug.org
\usepackage{verbatim}
\usepackage{amsmath}
\usepackage{hyperref}
\usepackage{enumitem}
\usepackage{setspace}
\usepackage[square,sort,comma,numbers]{natbib}

\title{Observation of magnetic field generation via the Weibel instability in interpenetrating plasma flows}
\date{}

\begin{document}

\maketitle

\noindent \author{C. M. Huntington$^{1}$\footnote{Author to whom correspondence should be addressed: huntington4@llnl.gov}, F. Fiuza$^{1}$, J. S. Ross$^{1}$, A. B. Zylstra$^{2}$, R. P. Drake$^{3}$, D. H. Froula$^{4}$, G. Gregori$^{5}$, N. L. Kugland$^6$, C. C. Kuranz$^{3}$, M. C. Levy$^{1}$, C. K. Li$^{2}$, J. Meinecke$^{5}$, T. Morita$^{7}$, R. Petrasso$^{2}$, C. Plechaty$^{1}$, B. A. Remington$^{1}$, D. D. Ryutov$^{1}$, Y. Sakawa$^{7}$, A. Spitkovsky$^{8}$, H. Takabe$^{7}$, H.-S. Park$^{1}$} \\

\begin{enumerate}[itemsep=0mm]
\item {\small \textit{Lawrence Livermore National Laboratory, Livermore, California 94550, USA}} %1
\item {\small \textit{Massachusetts Institute of Technology, Cambridge, Massachusetts 02139, USA}} %2
\item {\small \textit{Department of Atmospheric, Oceanic, and Space Sciences, University of Michigan, Ann Arbor, Michigan 48109, USA}} %3
\item {\small \textit{Physics Department and Laboratory for Laser Energetics, University of Rochester, Rochester, New York 14636, USA }}%4
\item {\small \textit{Department of Physics, University of Oxford, Parks Road, Oxford OX1 3PU, UK}} %5
\item {\small \textit{Lam Research Corporation, 4400 Cushing Parkway, Fremont, California 94538, USA}} %6
\item {\small \textit{Institute of Laser Engineering, Osaka University, Osaka 565-0871, Japan}} %7
\item {\small \textit{Department of Astrophysical Sciences, Princeton University, Princeton, New Jersey 08544, USA}} %8
\end{enumerate}

Collisionless shocks can be produced as a result of strong magnetic fields in a plasma flow, and therefore are common in many astrophysical systems.  The Weibel instability is one candidate mechanism for the generation of sufficiently strong fields to sustain a collisionless shock.  Despite their crucial role in astrophysical systems, observation of the magnetic fields produced by Weibel instabilities in experiments has been challenging.  Using a proton probe to directly image electromagnetic fields, we present evidence of Weibel-generated magnetic fields that grow in opposing, initially unmagnetized plasma flows from laser-driven laboratory experiments.  Three-dimensional particle-in-cell simulations reveal that the instability efficiently extracts energy from the plasma flows, and that the self-generated magnetic energy reaches a few percent of the total energy in the system. This result demonstrates an experimental platform suitable for the investigation of a wide range of astrophysical phenomena including collisionless shock formation in supernova remnants, large-scale magnetic field amplification, and the radiation signature from gamma-ray bursts.

The magnetic fields required for collisionless shock formation in astrophysical systems may either be initially present, for example in supernova remnants or young galaxies \cite{Bernet:2008}, or they may be self-generated in systems like gamma-ray bursts (GRBs) \cite{medvedevKorean}.  In the case of GRB outflows, the intense magnetic fields are greater than those which can be seeded by the GRB progenitor or produced by misaligned density and temperature gradients (the Biermann-battery effect) \cite{Medvedev:1999, Gruzinov:1999}. It has long been known that instabilities in can generate strong magnetic fields, even in the absence of seed fields. Weibel considered the development of an electromagnetic instability driven by the electron velocity anisotropy in a background of resting ions \cite{PhysRevLett.2.83}. The signature of the instability is a pattern of current filaments stretched along the axis of symmetry of the electron motion.  The process is quite general, and subsequent work has shown that such instabilities can be excited in both non-relativistic and relativistic shocks.  This general nature makes the Weibel instability common in astrophysical systems \cite{Moiseev:1963, Lazar:2009, Medvedev:2006}.  The instability provides a mechanism by which the electromagnetic turbulence associated with the formation of collisionless shocks is fed by the flow anisotropy of the protons (and ions) stochastically reflecting off of the shock \cite{Kato:2008, Spitkovsky:2008}, and leading ultimately to strong particle acceleration in GRB's \cite{Spitkovsky:2008a}.

The importance of  Weibel instabilities in astrophysical systems makes laboratory experiments that can access the collisionless plasma regime particularly compelling, though to date experiments have been limited to collisional systems (where Weibel instability growth is limited by collisional dissipation \cite{Ryutov:2014}) or those where the initial plasma conditions are not well characterized \cite{kuglands:natPhys2012, Fox:2013}.  Reaching the collisionless regime also permits the instability dynamics to be described by dimensionless parameters and scaled between laboratory and astrophysical systems \cite{Ryutov:2012}.  In the collisionless regime, the scaling is remarkably simple and allows one to predict the parameters of the unstable modes and the shocks (should they be formed) on the basis of laboratory measurements and the astrophysical ``input'' parameters, the density and velocity of the flows. 

\begin{figure}[ht]
\begin{center}
\includegraphics[width=4in]{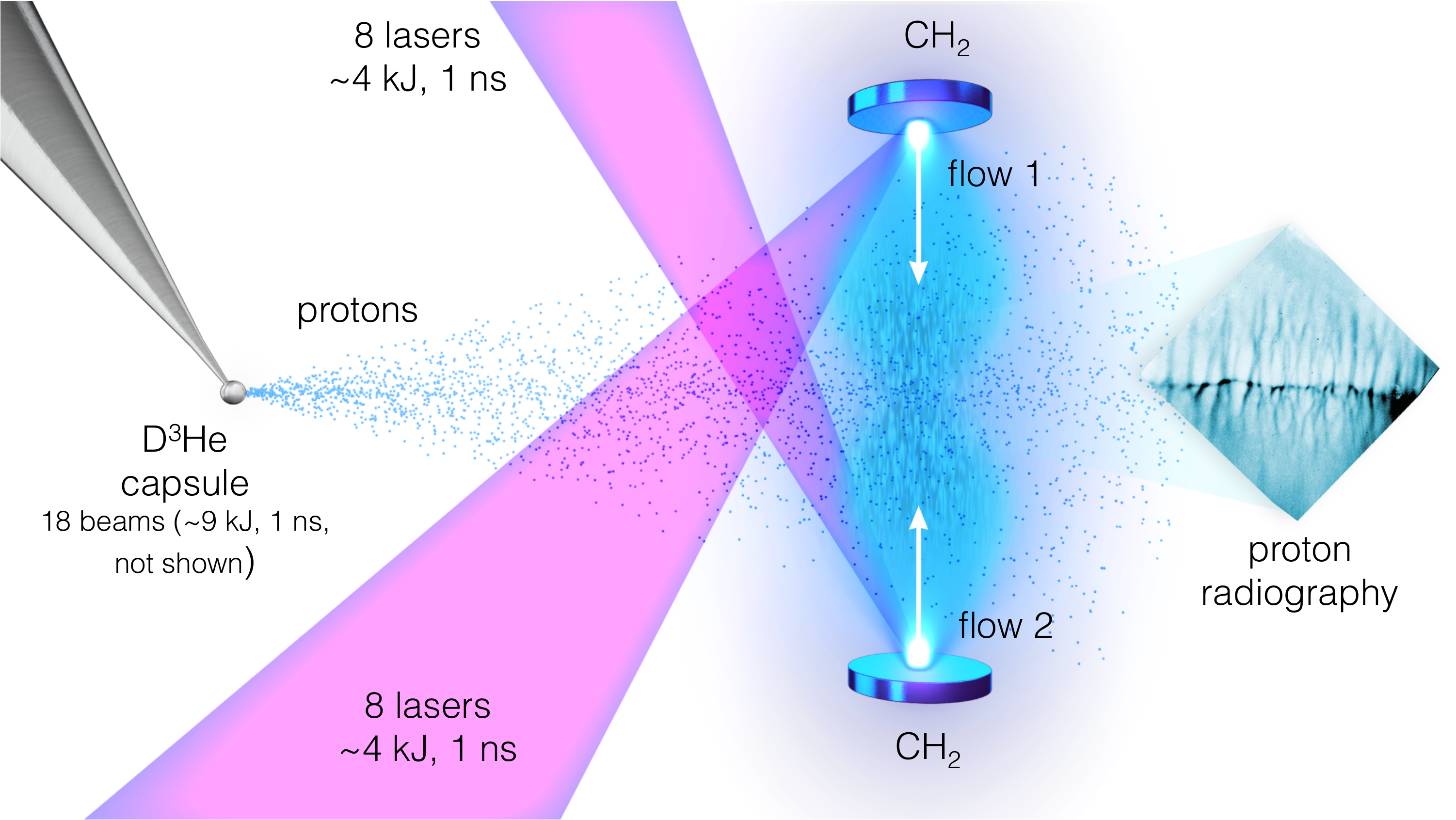}
\caption{\textbf{Experimental configuration to generate opposing plasma flows probed by D$^3$He protons.} The experiment consists of a pair of (CH$_2$) plastic foils of diameter 2 mm and thickness 500 $\mu$m, oriented face-on and separated by 8 mm.  Each was irradiated with 8 overlapped laser beams, delivering $\approx$4 kJ of 351 nm laser energy in a 1 ns square pulse. Distributed phase plates were used to produce super-Gaussian laser spots with focal spot diameters of 250 $\mu$m on the target surface.  After a delay, the proton probe was created by laser-compressing a thin-walled SiO$_2$ capsule. The capsule was filled with a 1:1 mixture of deuterium (D) and $^3$helium ($^3$He) at a total pressure of 18 atm.   At peak compression ($10^{23}$ cm$^{-3}$) protons are produced quasi-isotropically at energies of 3.0 and 14.7 MeV.   The protons were detected using a CR39 nuclear track detector positioned on the midplane of the CH$_2$ target foils, such that the protons traverse the central interaction region as shown.}
\label{fig:exp_setup}
\end{center}
\end{figure}

% Experiments
In experiments performed at the Omega Laser Facility \cite{Boehly:Omega}, we directly image the magnetic fields associated with the Weibel instability in well-characterized, counter-streaming plasma flows in the collisionless plasma regime \cite{ross:056501}.  The flows were established by laser-ablation of opposing foils, as in Fig. \ref{fig:exp_setup}.  The foils were oriented opposite each other and irradiated simultaneously, such that the expanding plasma flows interacted near the midplane between the foils \cite{kugland:056313, PhysRevLett.110.145005, Takabe:2008}. The plasma conditions in this geometry have been previously measured under identical conditions with Thomson scattering \cite{ross:056501}. When only a single foil was used, the plasma flow velocity $v$ was measured to be 1000 - 2000 km/s, with an electron density ($n_e) \approx 5\times$10$^{18}$ cm$^{-3}$ and an electron and ion temperature ($T_e$, $T_i$) less than 200 eV.  

When two opposing foils are used, as in the present work, the plasma density in  the counter-propagating flows grew by the anticipated factor of 2, while the electron and ion temperatures rapidly increased due to a combination of collisional electron heating and ion two-stream instability \cite{PhysRevLett.110.145005}. The ion instability quickly stabilizes as the electron and the ion temperatures equilibrate, which was observed to occur near 1 keV. The ions remain directed throughout the process, allowing competing instabilities, including the Weibel, to grow from the energy supplied by the flows \cite{Kato:2010}. 

\begin{figure}[ht]
\begin{center}
\includegraphics[width=4in]{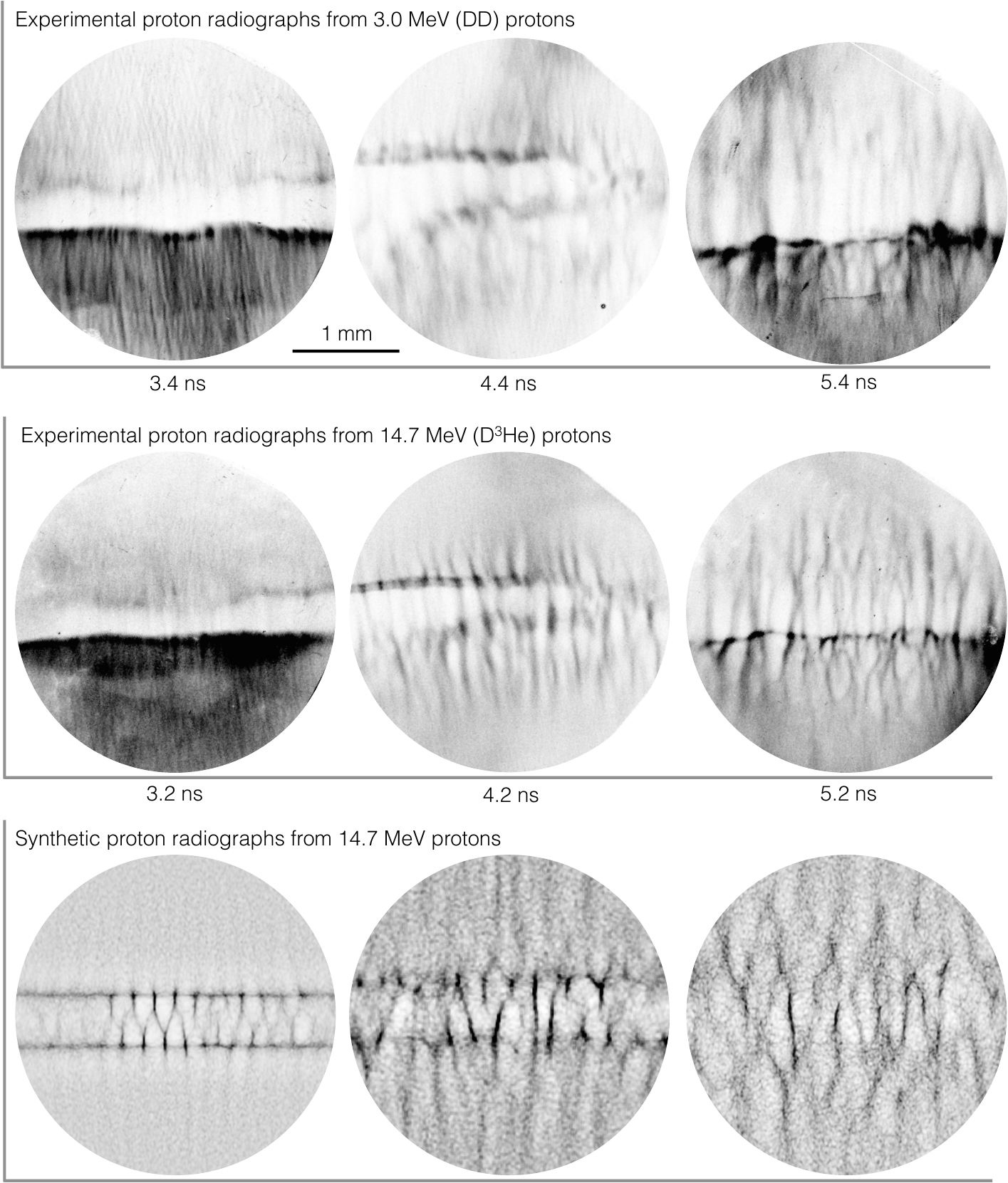}
\caption{\textbf{Experimental images and synthetic radiographs of magnetic field structures.} Experimental proton images are shown from 3.0 MeV DD protons (top row), 14.7 MeV D$^3$He protons (middle row), and synthetic 14.7 MeV proton tracing from 3D particle-in-cell (PIC) simulations (bottom row). In each case the plasma flows enter the frame from the top and bottom.  The small timing difference between DD and D$^3$He images is due to the proton time of flight from the capsule to the plasma interaction region.  At early time (approximately 3 ns after the drive begins), only initial traces of filamentation are observed.  At later times the filaments become more coherent and increase in extent along the flow direction.  In each case extended magnetic ``plates'' are formed above and below the midplane as a result of the large-scale Biermann battery fields generated in the laser ablation process \cite{Ryutov:2013}. All images are 3 mm in diameter in the target plane.}
\label{fig:data}
\end{center}
\end{figure}

In our experiment, the presence of magnetic fields is detected using proton imaging.  An isotropically-emitting proton source is generated by implosion of a capsule filled with D$^3$He, producing protons at 3 MeV (from D-D reactions) and at 14.7 MeV (from D - $^3$He reactions; see Supplementary Information for additional details on proton imaging). The protons that pass through the plasma interaction region are deflected by the electric and magnetic fields in the system, and are recorded using CR39 nuclear track detector at a magnification of $\approx30$. There are several important features in the proton radiography data, which was taken at three different times during the interaction of the flows, and is shown on the top two rows of Fig. \ref{fig:data}.  First, oriented along the flow direction is a pattern of filamentary structures, consistent with Weibel filamentation in the counter-propagating flows.  These features develop strongly between 3 and 4 ns after the start of the drive laser pulse and grow to lengths \textgreater1 mm along the direction of the flow.  The filamentary structure is clear in both 3.0 and 14.7 MeV proton images, and extend relatively uniformly for several mm across the plasma flow, occupying a total volume of several mm$^3$. The similarity in the observed features and relative contrast between the 3 MeV and 14.7 MeV radiographs indicate that proton deflections were produced by magnetic fields (see Supplementary Information for additional discussion).

In addition to the filaments, horizontal ``plate'' features are seen near the midplane of the drive plasmas.  These large-scale magnetic features have been observed in previous experiments with similar geometries \cite{kugland:056313, kuglands:natPhys2012}, and are understood to be the result of the initial Biermann battery-generated magnetic fields \cite{Ryutov:2013}. These fields are created at the target surface during the laser ablation and form a loop around the expanding plasma flow \cite{gregori:nature2012, Schoeffler:2013}.  The Biermann fields are frozen in the flow, following the effective electron trajectory to the midplane where the longitudinal electron velocity from the two flows is cancelled. The magnetic fields cannot readily cross the midplane and expand transversely, leading to the formation of characteristic plates \cite{Ryutov:2013}.  Asymmetry between the top and bottom plates in the data is related to slight non-uniformities in the flows, including differences in laser energy deposition on the two foils and tilts in their orientation relative to the proton probe. 

In order to better understand both the Weibel and Biermann battery-generated magnetic fields in the experiment we have conducted detailed 3-dimensional particle-in-cell (3D PIC) simulations with the code OSIRIS \cite{Fonseca:2002, Fonseca:2008} to model, from first principles, the counter-streaming plasma flows and the generation of electromagnetic fields (Fig \ref{fig:3dsim} a-b). The flows are initialized with the properties measured experimentally in the midplane region, namely each flow has n$_e$ = 5$\times 10^{18}$ cm$^{-3}$, $v$ = 1900 km/s, and T$_e$ = T$_i$ = 1 keV.  To include the effect of the Biermann battery, the flows were encircled by a large-scale magnetic field consistent with the misaligned density and temperature gradients of the flow, with an initial peak amplitude of 50 kGauss (see Ref. \cite{kugland:056313}).  Additional simulation details are found in the Supplementary Information section.

Within 1 ns of the opposing flows beginning to interact at the midpoint of the simulation volume, magnetic filaments are generated via the Weibel instability (Fig. \ref{fig:3dsim} b).  Additionally, the magnitude of the toroidal magnetic field related to the pre-imposed field doubles due to the conservation of magnetic flux. These fields lead to a long-range order in the system, and generate a pair of magnetic plates similar to those seen in the experiment (Fig. \ref{fig:data}). The presence of the toroidal fields does not significantly alter the formation of the ion Weibel instability, because the ions remain unmagnetized. This is supported by simulations where, when the initial magnetic fields are not included, Weibel filaments are still produced with the same structure.

To properly compare the PIC results with the experiment radiographs, we have simulated the proton imaging directly in the 3D OSIRIS simulations, to generate proton images of the electric and magnetic fields in the system. We assume an isotropic point source of 14.7 MeV protons located 1 cm from the center of the simulation box. The diagnostic protons interact with the 3-dimensional field structure during the simulation, and then are ballistically propagated to a 13 cm $\times$ 13 cm detector plane 30 cm from the source, matched to the imaging system in the experiment. 

The simulated proton radiographs are shown on the bottom row of Figure \ref{fig:data}, at the same times as the experimental data. To quantify the evolving structure in the system the filament spacing was measured for all images.  Shown in Fig. \ref{fig:3dsim} d), the growth in the size of the Weibel features is seen to be nearly equal for the measured and simulated images.  The increasing filament size indicates growing Weibel fields, and the efficiency of the instability to convert system kinetic energy into magnetic energy can be assessed directly from the simulations.  The magnetic energy associated with the instability is driven by the ion flows, and goes mainly into the transverse component of the field. The amplitude of the Weibel magnetic fields grows exponentially during the linear phase, with a growth rate of $\sim0.2~v/c \times \omega_{pi}$ (where $c$ is the speed of light and $\omega_{pi}$ is the ion plasma frequency), which is consistent with the linear theory of the instability. The linear phase of the instability saturates after 1 - 1.5 ns of interaction between the flows (i.e. after $\sim$2-3 e-foldings), though the field amplitude and filament size continue to increase in the subsequent nonlinear phase.  The significance of the magnetic energy in the system is quantified by the magnetization parameter $\sigma$, defined as $\sigma \simeq \left<B^2\right> / 4\pi m_i n_i v^2$, where the spatially-averaged field is given by $\left<B^2\right>$, and $m_i$ and $n_i$ represent the ion mass and density, respectively.  This ratio of magnetic energy to flow kinetic energy reaches nearly 1\% by the end of the experimental interaction time (approximately 5 ns in Fig. \ref{fig:3dsim} c).  

At times later than those probed in the experiment, the local magnetic field strength peaks at 0.6 MG, which corresponds to $\sigma =  5$\%. These high values illustrate the efficiency of the Weibel instability in converting kinetic energy into electromagnetic energy. At this amplitude the magnetic fields are large enough to cause the deflection of the incoming flows and the randomization of their kinetic energy. Furthermore, Weibel-mediated collisionless shocks are believed to occur at this level of magnetization, provided that there is a  sufficiently large interpenetration distance, of the order of 300 $c/\omega_{pi}$ \cite{Kato:2008}. This condition precludes shock formation in the present experiment, where this length is only 55 $c/\omega_{pi}$; however it should be observed with a similar setup if higher densities and/or longer flows are generated.

\begin{figure}[ht]
\begin{center}
\includegraphics[width=5.0in]{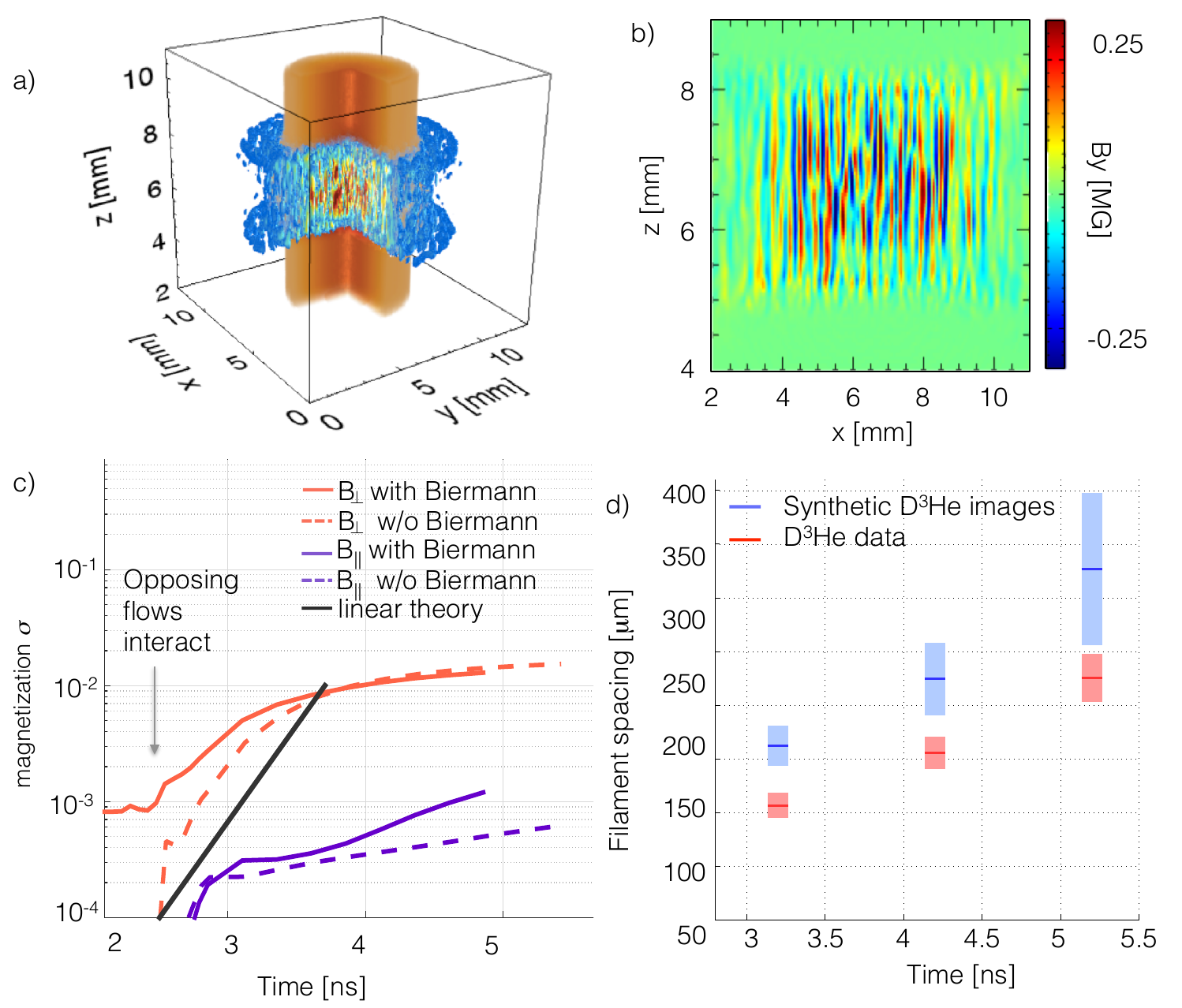}
\caption{\textbf{Temporal evolution of magnetic field magnitude from simulation and field structure from experimental images.}  a) 3D OSIRIS simulation of the system after 1 ns of interaction between the counter-streaming 1900 km/s plasma flows (approximately 3 ns after the experimental drive laser pulse; flows enter from top and bottom). Magnetic fields are shown qualitatively in the blue/red color scale, with electron density in orange. b) Magnetic field slice (transverse magnetic field component B$_y$) along the $y$-axis midplane, at the same time, illustrating the presence of strong filaments associated with the Weibel instability.  c) Plasma magnetization $\sigma$ as a function of time. When the flows are initiated with zero initial magnetic field (dashed lines) the magnetizations remains at zero until the flows begin interacting, between 2 and 3 ns.  When initial toroidal fields are included consistent with the Biermann battery mechanism, the perpendicular magnetization is $\sim0.1\%$ before the flows interact (solid colored lines).  In both cases the magnetic energy associated with Weibel instability increases sharply after the flows interact, increasing $\sigma$ by a factor greater than 10 in several ns.  The magnetization due to the ion Weibel instability, growing at the theoretical linear growth rate is shown in solid black.  This calculation shows that the Weibel-generated magnetization becomes the dominant contribution to the overall magnetization of the system.  d) Measurement of the mean separation between filaments in experimental proton radiographs (red) and synthetic proton images from 3D PIC simulations (blue).  The filament spacing approximately doubles over the 2 ns of observation. Note that time is experimental time, measured with respect to the beginning of the drive laser.} 
\label{fig:3dsim}
\end{center}
\end{figure}

The dimensionless magnetization parameter $\sigma$ serves to connect experiments to astrophysical systems, and can be directly applied to scaled systems of interest \cite{Ryutov:2012}.  For example, multiwavelength observations of afterglow emission of GRBs suggest sub-equipartition levels of magnetic field in the region behind the forward shock ($\sigma \approx 0.01 - 0.1$) \cite{Frail:2000}. This field is likely generated near a collisionless shock front, as the pre-shock interstellar medium is effectively unmagnetized ($\sigma \approx 10^{-10}$). Our experiments imply that even non-relativistic Weibel instability in an initially unmagnetized medium is capable of generating the percent-level magnetization observed in GRBs; collisions of relativistic flows are likely to produce even stronger fields. \\

\section*{Supplementary Information}
% http://www.nature.com/nphys/authors/submit/index.html#Supplementary-information

% \begin{comment}
\subsection*{OMEGA Experimental Details} 
The experimental geometry consists of a pair of polyethylene (CH$_2$) plastic foils of diameter 2 mm and thickness 500 $\mu$m were oriented face-on, separated by 8 mm.  Each was irradiated with 8 overlapped laser beams from the OMEGA laser, delivering  $\approx$4 kJ of 351 nm laser energy in a 1 ns square pulse. Distributed phase plates were used to produce super-Gaussian laser spots with focal spot diameters of 250 $\mu$m on the target surface.  The expanding plasma plumes interact at the midplane between the targets.  

After a delay of 3 - 5 ns from the beginning of the drive pulse,  the proton probe was created by compressing a thin-walled SiO$_2$ capsule with 18 beams, delivering $\approx$9 kJ total laser energy. The capsule was filled with a 1:1 mixture of deuterium (D) and $^3$helium ($^3$He) at a total pressure of 18 atm.   At peak compression ($10^{23}$ cm$^{-3}$), protons are produced quasi-isotropically at 3.0 MeV through DD reactions, and at 14.7 MeV through fusion of D and $^3$He  \cite{manuel:063506, li:10E725}.  The details of proton imaging have been treated at length in literature (see \cite{kugland:101301} and references therein), and proton probing has been used in numerous high-energy-density experiments on OMEGA and elsewhere to image electric and magnetic field structures (See \cite{Zylstra:2012} and references therein).  The protons were detected using CR39 nuclear track detector positioned on the midplane of the CH$_2$ target foils, such that the protons traverse the central interaction region as shown in Fig. 1 of the main text. 

\subsection*{3D OSIRIS Simulations}

The PIC simulations presented here were done with the fully electromagnetic, fully relativistic, and massively parallel code OSIRIS \cite{Fonseca:2002, Fonseca:2008}. The code solves Maxwell's equations directly, resolving all the relevant physics at the electron and ion skin depth scales. The relativistic Lorentz force is used to calculate the motion of the plasma particles, and relativistic expressions are used to derive the charge and current densities from the positions and momenta of the particles. Plasma electromagnetic and electrostatic instabilities arise in the simulations from first principles, as the simulations use a fully kinetic model for the plasma particles.

The simulation used to study the interaction between counter-streaming plasma flows has a box size of 1.3 cm (90 c/$\omega_{pi}$) in each direction and ran for a total of 6 ns ($\approx1.7\times10^4 \omega_{pi}^{-1}$). Each plasma flow is composed by an electron and ion species (assumed to be fully ionized and modeled with $m_i/(Z\times m_e)$ = 128). The numerical parameters were as follows: the 3D simulations used at least 2 cells per electron skin depth, 22 cells per ion skin depth, and 2 particles per cell per species, for a total of 70 billion particles. Due to the outstanding computational requirements, the 3D simulation ran in 131,072 cores in the supercomputer Mira (ANL). All simulations used cubic particle shapes, and current and field smoothing with compensation for improved numerical properties. Additional 2D simulations (not shown here) were done with higher resolution, greater number of particles per cell, and realistic ion to electron mass ratio ($m_i/(Z \times m_e$) = 2048), confirming overall result convergence consistent with the 3D results and showing that the ion Weibel instability can be reasonably scaled between systems with different mass ratios and (non-relativistic) flow velocities \cite{Ryutov:2012}. Additional detailed analysis of the simulations performed will be presented in a separate publication.

The simulated proton radiographs were obtained by launching a 14.7 MeV proton beam transversely to the flow propagation direction. The proton distribution was initialized in OSIRIS following the distribution of an isotropic point source located 1 cm away from the beginning of the simulation box, in order to be consistent with our experimental setup. The protons probe the self-consistent fields produced in the 3D simulation and exit on the opposite side of the simulation box, being then propagated ballistically to a square detector of 13 cm $\times$ 13 cm placed 30 cm away from the original point source, matching the experimental magnification of 30$\times$. The detector has 512 $\times$ 512 points, and $\sim10$ million probing protons are collected in each image.

\subsection*{Interpretation of field structure from proton radiographs}
The interpretation of proton images from complex systems must take into account the susceptibility of protons to deflections by both electric and magnetic fields.  One can break the degeneracy between $E$ and $B$ fields is by comparing the relative deflection of higher and lower energy protons \cite{0741-3335-52-12-124027,Li:2009, kugland:101301}.  The distinct proton populations produced by the D$^3$He implosion lend themselves to this method. For the respective fields $E$ and $B$, the particle deflection $\sigma$ is given by: 
\begin{align}
\theta_B = \frac{q}{\sqrt{2m_pE_p}}\int{B_{\perp}dl}\\
\theta_E = \frac{q}{2E_p}\int{E_{\perp}dl}.
\end{align}
Thus, the ratio of deflection for 14.7 and 3 MeV protons expected from B-fields is $\theta_{DD} / \theta_{D^3He} \propto \sqrt{14.7/3} \approx 2.2$, while from E-fields one expects $\theta_{DD}/\theta_{D^3He} \propto 14.7/3 \approx 4.9$. 

While the ratios above could in principle be directly measured, the complex, 3D structure of the system under investigation makes a quantitative comparison between low and high-energy proton images difficult.  However, the similarity between the images from 3.0 and 14.7 MeV protons suggests deflection of the protons by magnetic fields.  In particular, the same filaments can be co-registered between the two images, at decreased contrast in the image from 3.0 MeV protons.  This is consistent with deflection from magnetic fields; were the deflection of the lower-energy protons $4.9\times$ greater than the high-energy particles, the protons deflected by the small-scale filaments that are clear in the D$^3$He image would be more diffuse in the DD image.  Similarly, were the horizontal ``plates'' the result of electric fields, the difference in the position and contrast between the two proton energies would be larger, closer to 4.9$\times$, which is not seen in the data.

The implementation of the toroidal Biermann-like fields into the 3D PIC simulations has been described in the main text; we comment here on the effect of these fields on the Weibel filaments that are the focus of this work.  It is important to note that in the experiment, the Biermann battery and Weibel-generated fields are effectively independent of each other.  The Biermann fields are the result of gradients in density and temperature ($\dot{B}\propto\nabla T_e\times\nabla n_e$), which arise naturally in the ablated plasma flows.  However, these fields are strongest near the surface of the targets where the transverse gradients are largest, and are zero on-axis, where the Weibel instability mediated fields are strongest.  In addition to this spatial separation, the Biermann battery and Weibel modes are also clearly separated in $k$-space, inhibiting efficient coupling between the two.  The presence of the large-scale structure (the horizontal ``plates'') in the proton radiographs is related to this difference in scales--the large spatial extent of the Biermann fields generates a sizable proton deflection (related to $\int B\cdot dL$), despite their relatively low field strength.

\subsection*{Analytic treatment of growing modes}

To assess the susceptibility of the plasma in our experiment to Weibel growth, we have performed a linear stability analysis based on the collisionless Vlasov equation \cite{Kato:2010}. Using the same techniques as previous studies \cite{davidson:317, berger:3}, we arrived at the dispersion relation properly accounting for the chemical composition of the target. Such a description is necessary for multi-species plasmas, including the present system consisting of carbon and hydrogen. The results of this analysis show that the linear growth rate for a plasma with temperature of $\sim$1 keV, as measured in the system of interest, is sufficient for Wiebel filaments to reach a well-developed state during the first 1-2 ns of interaction between the plasma flows (3 - 4 ns after the initial laser drive).  This agrees well with the growth observed in both experiment and simulation.  

The linear dispersion relation for the filamentation instability driven by counter-streaming ion flows has been considered in a number of papers. The most relevant for our analysis are Refs.\cite{davidson:317, berger:3}. In our experiment, we need to consistently account for the presence of multiple ion species; the presence of the light ions leads to an enhancement in the stabilizing effect of a finite ion temperature. For the electromagnetic Weibel mode propagating perpendicularly with respect to the flow direction, the dispersion relation reads as:
\begin{align}
k^2_x c^2 + \frac{\omega_{pe}^2}{1 + \frac{\left|k\right|}{\Gamma}\sqrt{\frac{2T_e}{\pi m_e}}} + \omega_{pi}^2 \sum{C_\alpha\left[G_1\left(\frac{\Gamma^2 A_\alpha m_p}{2k^2T_\alpha}\right) - \frac{k^2_xv^2}{\Gamma^2}G_2\left(\frac{\Gamma^2 A_\alpha m_p}{2k^2T_\alpha}\right)\right]}= 0,
\end{align}
Here $c$ is the speed of light and $k$ is the wave number perpendicular to the flow direction. The flow velocity is $v$, subscript ``e'' refers to the electron parameters, the subscript ``$\alpha$'' refers to the parameters of a certain ion species. We consider symmetric flows for which the unstable mode is the mode of an exponential growth.  It is assumed that the electron thermal velocity exceeds the flow velocity, a condition that usually holds by a very large margin both in the laboratory and non-relativistic astrophysics. The growth rate is denoted by $\Gamma$, and $G_1$ and $G_2$  are dimensionless functions defined for $y>0$ as 
\begin{align}
G_1(y) = \frac{1}{\sqrt{\pi}}\int^{+\infty}_{-\infty}\frac{ye^{-x^2}}{x^2+y} dx ;\\ 
G_2(y) = \frac{2y}{\sqrt{\pi}}\int^{+\infty}_{-\infty}\frac{x^2e^{-x^2}}{x^2+y} dx.
\end{align}
Additionally, one has
\begin{align}
\omega^2_{pi} &= \frac{4\pi e^2}{m_p}\sum\limits_{\alpha}{\frac{Z_\alpha^2}{A_\alpha}n_\alpha}, \\ 
\omega^2_{pe} &= \frac{4\pi e^2}{m_e}\sum\limits_{\alpha}{n_\alpha}, \text{ and} \\ 
C_Z &= \frac{n_\alpha Z_\alpha^2/A_\alpha}{\sum\limits_{\alpha}{n_\alpha Z_\alpha^2/A_\alpha}}.
\end{align}
Here $m_e$ and $m_p$ are the electron and proton mass, $Z_\alpha$, $A_\alpha$ and $n_\alpha$ are the charge, the atomic number and the particle density of the ion species $\alpha$. 

By introducing dimensionless units for the wave number and growth rate,
\begin{align}
\tilde{\Gamma} = \Gamma \frac{c}{v\omega_{pi}} \\
\tilde{k} = k_x\frac{c}{\omega_{pi}}.
\end{align}
one can recast the dispersion relation to the dimensionless form:
\begin{align}
\tilde{k}^2 + \frac{a_1} {1+\sqrt{\frac{a_2}{\pi}}\frac{\left|\tilde{k}\right|}{\tilde{\Gamma}}} + \sum{C_Z\left[G_1\left(\frac{a_{3\alpha}\tilde{\Gamma}^2}{\tilde{k}^2}\right) - \frac{\tilde{k}^2}{\tilde{\Gamma}^2} G_2\left(\frac{a_{3\alpha}\tilde{\Gamma}^2}{\tilde{k}^2}\right) \right]} = 0,
\label{eqn:disp}
\end{align}
where
\begin{align}
a_1 &= \frac{\omega_{pe}^2}{\omega_{pi}^2}, \\
a_2 &= \frac{2T_e}{m_ev^2}, ~~\text{and} \\
a_{3\alpha} &= \frac{2T_\alpha}{A_\alpha m_pv^2}.
\end{align}
This form allows one to more readily compare the results of simulations and experiments between systems, including astrophysical systems. The characteristic dispersion curves for the conditions of the OMEGA experiment are shown in Fig. \ref{fig:3dsim}. 

\begin{figure}[ht]
\begin{center}
\includegraphics[width=3.0in]{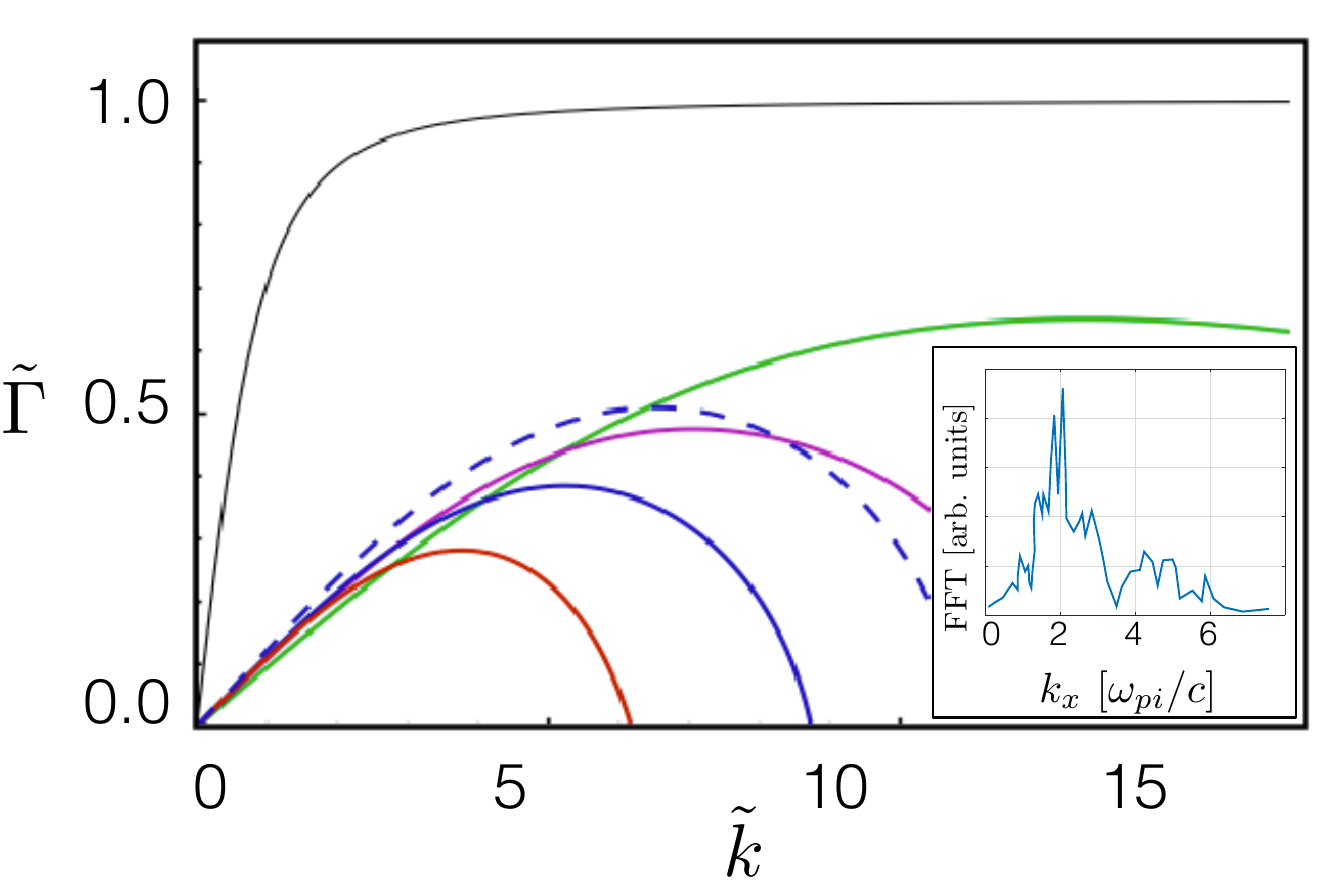}
\caption{The linear Weibel growth rate $\tilde\Gamma$ vs the wave number $\tilde k$. The green, magenta, blue, and red curves correspond to CH$_2$ flows at electron and ion temperatures of 0.1 keV (green), 0.5 keV (magenta), 1 keV (blue) and 2 keV (red). The maximum growth rate for the electron density of 10$^{19}$  cm$^{-3}$ in the CH$_2$ plasma is $0.5\times10^{10}~s^{-1}$ for blue curve. The dashed blue curve is for pure carbon at $T_e=T_i=1$ keV, so that the difference between the solid and dashed blue curves is a manifestation of stabilization by the light ions.  The black curve is a reference growth rate $\Gamma=k$v$\omega_{pi} / \sqrt{k^2c^2 + \omega^2_{pi}}$.  Finally, the inset plot shows the magnetic field mode distribution from simulations.  Here $k_x$ is transverse to the flow and measured after 2 ns of flow interaction, showing a range of unstable modes observed consistent with theoretical analysis.}
\label{fig:3dsim}
\end{center}
\end{figure}

\clearpage

\noindent \textbf{Correspondence:} 
Correspondence and requests for materials should be addressed to C. M. Huntington.~(email: huntington4@llnl.gov). \\

\noindent \textbf{Acknowledgements:}
\noindent We thank the staff of the Omega Laser Facility for their experimental support. This work was performed under the auspices of the US Department of Energy by the Lawrence Livermore National Laboratory, under Contract No. DE-AC52-07NA27344, with funding support from LLNL LDRD grant No. 11-ERD-054 and from the European Research Council under the European Community's Seventh Framework Programme (FP7/2007-2013), ERC grant agreement no. 256973. Computing support for this work came from ALCC and INCITE awards on Mira (ALCF supported under contract DE-AC02-06CH11357) and from the LLNL Institutional Computing Grand Challenge program on Vulcan.  Additionally, the authors would like to acknowledge the OSIRIS Consortium, consisting of UCLA and IST (Lisbon, Portugal) for the use of  the OSIRIS 2.0 framework and the visXD framework.  F.F. acknowledges the LLNL Lawrence Fellowship for financial support. A.S. is supported by DOE grant DE-NA0002200. \\
% \item[Competing Interests] The authors declare that they have no competing financial interests.
% \end{addendum}

% \noindent \author{C. M. Huntington F. Fiuza, J. S. Ross, A. B. Zylstra, R. P. Drake, D. H. Froula, G. Gregori, N. L. Kugland, C. C. Kuranz, M. C. Levy, C. K. Li, J. Meinecke, T. Morita, R. Petrasso, C. Plechaty, B. A. Remington, D. D. Ryutov, Y. Sakawa, A. Spitkovsky, H. Takabe, H.-S. Park 

%\noindent \textbf{Authors Contribution:}
%This experimental platform was conceived by G.G., B.A.R., R.P.D., H.-S.P., D.H.F., Y.S., A.S., and H.T..  It was developed in a series of initial experiments by N.L.K., C.C.K., J.M., T.M, C.P., J.S.R., and C.M.H., and the data shown here were collected in experiments by C.M.H., J.S.R., and H.-S.P..  Theoretical support for this work was provided by A.S., G.G., and primarily by F.F., and D.R..   Support for proton diagnostics was provided by C.K.L., R.P., and particularly A.B.Z.. Particle-in-cell simulations were performed and synthetic radiographs produced by F.F..  Additional contributions to the analysis of proton radiographs by H.-S.P., J.S.R., M.C.L., A.B.Z., D.R., A.S., and C.M.H..  The manuscript was written by C.M.H., F.F., J.S.R., and D.R.. \\

% \noindent Supplementary Information is linked to the online version of the paper at www.nature.com/nature.

\bibliographystyle{unsrt}
% \bibliography{$HOME/Documents/Papers/masterBib} 

\begin{thebibliography}{10}

\bibitem{Bernet:2008}
Martin~L. Bernet, Francesco Miniati, Simon~J. Lilly, Philipp~P. Kronberg, and
  Miroslava Dessauges-Zavadsky.
\newblock Strong magnetic fields in normal galaxies at high redshift.
\newblock {\em Nature}, 454(7202):302--304, 07 2008.

\bibitem{medvedevKorean}
Mikhail~V. Medvedev, Luis~O. Silva, Massimiliano Fiore, Ricardo~A. Fonseca, and
  Warren~B. Mori.
\newblock Generation of magnetic fields in cosmological shocks.
\newblock {\em Journal of the Korean astronomical society}, 37(5):533--541,
  2004.

\bibitem{Medvedev:1999}
Mikhail~V. Medvedev and Abraham Loeb.
\newblock Generation of magnetic fields in the relativistic shock of gamma-ray
  burst sources.
\newblock {\em The Astrophysical Journal}, 526(2):697, 1999.

\bibitem{Gruzinov:1999}
A.~Gruzinov and E.~Waxman.
\newblock Gamma-ray burst afterglow: Polarization and analytic light curves.
\newblock {\em Astrophys. J.}, 511:852--861, Feb 1999.

\bibitem{PhysRevLett.2.83}
Erich~S. Weibel.
\newblock Spontaneously growing transverse waves in a plasma due to an
  anisotropic velocity distribution.
\newblock {\em Phys. Rev. Lett.}, 2:83--84, Feb 1959.

\bibitem{Moiseev:1963}
S~S Moiseev and R~Z Sagdeev.
\newblock Collisionless shock waves in a plasma in a weak magnetic field.
\newblock {\em Journal of Nuclear Energy. Part C, Plasma Physics, Accelerators,
  Thermonuclear Research}, 5(1):43, 1963.

\bibitem{Lazar:2009}
M.~{Lazar}, R.~{Schlickeiser}, R.~{Wielebinski}, and S.~{Poedts}.
\newblock {Cosmological Effects of Weibel-Type Instabilities}.
\newblock {\em The Astrophysical Journal}, 693:1133--1141, March 2009.

\bibitem{Medvedev:2006}
Mikhail~V. Medvedev.
\newblock Electron acceleration in relativistic gamma-ray burst shocks.
\newblock {\em The Astrophysical Journal Letters}, 651(1):L9, 2006.

\bibitem{Kato:2008}
Tsunehiko~N. Kato and Hideaki Takabe.
\newblock Nonrelativistic collisionless shocks in unmagnetized electron-ion
  plasmas.
\newblock {\em The Astrophysical Journal Letters}, 681(2):L93, 2008.

\bibitem{Spitkovsky:2008}
Anatoly Spitkovsky.
\newblock On the structure of relativistic collisionless shocks in electron-ion
  plasmas.
\newblock {\em The Astrophysical Journal Letters}, 673(1):L39, 2008.

\bibitem{Spitkovsky:2008a}
Anatoly Spitkovsky.
\newblock Particle acceleration in relativistic collisionless shocks: Fermi
  process at last?
\newblock {\em The Astrophysical Journal Letters}, 682(1):L5, 2008.

\bibitem{Ryutov:2014}
D.~D. Ryutov, F.~Fiuza, C.~M. Huntington, J.~S. Ross, and H.-S. Park.
\newblock Collisional effects in the ion weibel instability for two
  counter-propagating plasma streams.
\newblock {\em Physics of Plasmas}, 21(3), 2014.

\bibitem{kuglands:natPhys2012}
N.~L. Kugland, D.~D. Ryutov, P-Y. Chang, R.~P. Drake, G.~Fiksel, D.~H. Froula,
  S.~H. Glenzer, G.~Gregori, M.~Grosskopf, M.~Koenig, Y.~Kuramitsu, C.~Kuranz,
  M.~C. Levy, E.~Liang, J.~Meinecke, F.~Miniati, T.~Morita, A.~Pelka,
  C.~Plechaty, R.~Presura, A.~Ravasio, B.~A. Remington, B.~Reville, J.~S. Ross,
  Y.~Sakawa, A.~Spitkovsky, H.~Takabe, and H-S. Park.
\newblock Self-organized electromagnetic field structures in laser-produced
  counter-streaming plasmas.
\newblock {\em Nature Physics}, 8(11):809--812, 2012.

\bibitem{Fox:2013}
W.~Fox, G.~Fiksel, A.~Bhattacharjee, P.-Y. Chang, K.~Germaschewski, S.~X. Hu,
  and P.~M. Nilson.
\newblock Filamentation instability of counterstreaming laser-driven plasmas.
\newblock {\em Phys. Rev. Lett.}, 111:225002, Nov 2013.

\bibitem{Ryutov:2012}
D~D Ryutov, N~L Kugland, H~S Park, C~Plechaty, B~A Remington, and J~S Ross.
\newblock Basic scalings for collisionless-shock experiments in a plasma
  without pre-imposed magnetic field.
\newblock {\em Plasma Physics and Controlled Fusion}, 54(10):105021, 2012.

\bibitem{Boehly:Omega}
T.~R. Boehly, R.~S. Craxton, T.~H. Hinterman, J.~H. Kelly, T.~J. Kessler, S.~A.
  Kumpan, S.~A. Letzring, R.~L. McCrory, S.~F.~B. Morse, W.~Seka, S.~Skupsky,
  J.~M. Soures, and C.~P. Verdon.
\newblock The upgrade to the omega laser system.
\newblock {\em Review of Scientific Instruments}, 66(1):508--510, 1995.

\bibitem{ross:056501}
J.~S. Ross, S.~H. Glenzer, P.~Amendt, R.~Berger, L.~Divol, N.~L. Kugland, O.~L.
  Landen, C.~Plechaty, B.~Remington, D.~Ryutov, W.~Rozmus, D.~H. Froula,
  G.~Fiksel, C.~Sorce, Y.~Kuramitsu, T.~Morita, Y.~Sakawa, H.~Takabe, R.~P.
  Drake, M.~Grosskopf, C.~Kuranz, G.~Gregori, J.~Meinecke, C.~D. Murphy,
  M.~Koenig, A.~Pelka, A.~Ravasio, T.~Vinci, E.~Liang, R.~Presura,
  A.~Spitkovsky, F.~Miniati, and H.-S. Park.
\newblock Characterizing counter-streaming interpenetrating plasmas relevant to
  astrophysical collisionless shocks.
\newblock {\em Physics of Plasmas}, 19(5):056501, 2012.

\bibitem{kugland:056313}
N.~L. Kugland, J.~S. Ross, P.-Y. Chang, R.~P. Drake, G.~Fiksel, D.~H. Froula,
  S.~H. Glenzer, G.~Gregori, M.~Grosskopf, C.~Huntington, M.~Koenig,
  Y.~Kuramitsu, C.~Kuranz, M.~C. Levy, E.~Liang, D.~Martinez, J.~Meinecke,
  F.~Miniati, T.~Morita, A.~Pelka, C.~Plechaty, R.~Presura, A.~Ravasio, B.~A.
  Remington, B.~Reville, D.~D. Ryutov, Y.~Sakawa, A.~Spitkovsky, H.~Takabe, and
  H.-S. Park.
\newblock Visualizing electromagnetic fields in laser-produced
  counter-streaming plasma experiments for collisionless shock laboratory
  astrophysics.
\newblock {\em Physics of Plasmas}, 20(5):056313, 2013.

\bibitem{PhysRevLett.110.145005}
J.~S. Ross, H.-S. Park, R.~Berger, L.~Divol, N.~L. Kugland, W.~Rozmus,
  D.~Ryutov, and S.~H. Glenzer.
\newblock Collisionless coupling of ion and electron temperatures in
  counterstreaming plasma flows.
\newblock {\em Phys. Rev. Lett.}, 110:145005, Apr 2013.

\bibitem{Takabe:2008}
H~Takabe, T~N Kato, Y~Sakawa, Y~Kuramitsu, T~Morita, T~Kadono, K~Shigemori,
  K~Otani, H~Nagatomo, T~Norimatsu, S~Dono, T~Endo, K~Miyanishi, T~Kimura,
  A~Shiroshita, N~Ozaki, R~Kodama, S~Fujioka, H~Nishimura, D~Salzman,
  B~Loupias, C~Gregory, M~Koenig, J~N Waugh, N~C Woolsey, D~Kato, Y-T Li, Q-L
  Dong, S-J Wang, Y~Zhang, J~Zhao, F-L Wang, H-G Wei, J-R Shi, G~Zhao, J-Y
  Zhang, T-S Wen, W-H Zhang, X~Hu, S-Y Liu, Y~K Ding, L~Zhang, Y-J Tang, B-H
  Zhang, Z-J Zheng, Z-M Sheng, and J~Zhang.
\newblock High-mach number collisionless shock and photo-ionized non-lte plasma
  for laboratory astrophysics with intense lasers.
\newblock {\em Plasma Physics and Controlled Fusion}, 50(12):124057, 2008.

\bibitem{Kato:2010}
Tsunehiko~N. Kato and Hideaki Takabe.
\newblock Electrostatic and electromagnetic instabilities associated with
  electrostatic shocks: Two-dimensional particle-in-cell simulation.
\newblock {\em Physics of Plasmas}, 17(3):032114, 2010.

\bibitem{Ryutov:2013}
D.~D. Ryutov, N.~L. Kugland, M.~C. Levy, C.~Plechaty, J.~S. Ross, and H.~S.
  Park.
\newblock Magnetic field advection in two interpenetrating plasma streams.
\newblock {\em Physics of Plasmas}, 20(3):032703, 2013.

\bibitem{gregori:nature2012}
G~Gregori, A~Ravasio, C.~D Murphy, K~Schaar, A~Baird, A.~R Bell,
  A~Benuzzi-Mounaix, R~Bingham, C~Constantin, R.~P Drake, M~Edwards, E.~T
  Everson, C.~D Gregory, Y~Kuramitsu, W~Lau, J~Mithen, C~Niemann, H.-S Park,
  B.~A Remington, B~Reville, D.~D Robinson, A. P. L.~Ryutov, Y~Sakawa, S~Yang,
  N.~C Woolsey, and F~and Koenig, M.~Miniati.
\newblock Generation of scaled protogalactic seed magnetic fields in
  laser-produced shock waves.
\newblock {\em Nature Letters}, 481(11):480--483, 2012.

\bibitem{Schoeffler:2013}
K.~M. {Schoeffler}, N.~F. {Loureiro}, R.~A. {Fonseca}, and L.~O. {Silva}.
\newblock {Magnetic field generation and amplification in an expanding plasma}.
\newblock {\em ArXiv e-prints}, August 2013.

\bibitem{Fonseca:2002}
R.~Fonseca, L.~Silva, F.~Tsung, V.~Decyk, W.~Lu, C.~Ren, W.~Mori, S.~Deng,
  S.~Lee, T.~Katsouleas, and J.~Adam.
\newblock Osiris: A three-dimensional, fully relativistic particle in cell code
  for modeling plasma based accelerators.
\newblock In Peter Sloot, Alfons Hoekstra, C.~Tan, and Jack Dongarra, editors,
  {\em Computational Science, ICCS 2002}, volume 2331 of {\em Lecture Notes in
  Computer Science}, pages 342--351. Springer Berlin / Heidelberg, 2002.

\bibitem{Fonseca:2008}
R~A Fonseca, S~F Martins, L~O Silva, J~W Tonge, F~S Tsung, and W~B Mori.
\newblock One-to-one direct modeling of experiments and astrophysical
  scenarios: pushing the envelope on kinetic plasma simulations.
\newblock {\em Plasma Physics and Controlled Fusion}, 50(12):124034, 2008.

\bibitem{Frail:2000}
D.~A. Frail, E.~Waxman, and S.~R. Kulkarni.
\newblock A 450 day light curve of the radio afterglow of grb 970508: Fireball
  calorimetry.
\newblock {\em The Astrophysical Journal}, 537(1):191, 2000.


\bibitem{manuel:063506}
M.~J.-E. Manuel, A.~B. Zylstra, H.~G. Rinderknecht, D.~T. Casey, M.~J.
  Rosenberg, N.~Sinenian, C.~K. Li, J.~A. Frenje, F.~H. S\'{e}guin, and R.~D.
  Petrasso.
\newblock Source characterization and modeling development for
  monoenergetic-proton radiography experiments on omega.
\newblock {\em Review of Scientific Instruments}, 83(6):063506, 2012.

\bibitem{li:10E725}
C.~K. Li, F.~H. S\'{e}guin, J.~A. Frenje, {\em et al.}
\newblock Monoenergetic proton backlighter for measuring e and b fields and for
  radiographing implosions and high-energy density plasmas (invited).
\newblock {\em Review of Scientific Instruments}, 77(10):10E725, 2006.

\bibitem{kugland:101301}
N.~L. Kugland, D.~D. Ryutov, C.~Plechaty, J.~S. Ross, and H.-S. Park.
\newblock Invited article: Relation between electric and magnetic field
  structures and their proton-beam images.
\newblock {\em Review of Scientific Instruments}, 83(10):101301, 2012.

\bibitem{Zylstra:2012}
A.~B. Zylstra, C.~K. Li, H.~G. Rinderknecht, F.~H. S\'{e}guin, R.~D. Petrasso,
  C.~Stoeckl, D.~D. Meyerhofer, P.~Nilson, T.~C. Sangster, S.~Le Pape,
  A.~Mackinnon, and P.~Patel.
\newblock Using high-intensity laser-generated energetic protons to radiograph
  directly driven implosions.
\newblock {\em Review of Scientific Instruments}, 83(1):013511, 2012.

\bibitem{Fonseca:2002}
R.~Fonseca, L.~Silva, F.~Tsung, V.~Decyk, W.~Lu, C.~Ren, W.~Mori, S.~Deng,
  S.~Lee, T.~Katsouleas, and J.~Adam.
\newblock Osiris: A three-dimensional, fully relativistic particle in cell code
  for modeling plasma based accelerators.
\newblock In Peter Sloot, Alfons Hoekstra, C.~Tan, and Jack Dongarra, editors,
  {\em Computational Science, ICCS 2002}, volume 2331 of {\em Lecture Notes in
  Computer Science}, pages 342--351. Springer Berlin / Heidelberg, 2002.

\bibitem{Fonseca:2008}
R~A Fonseca, S~F Martins, L~O Silva, J~W Tonge, F~S Tsung, and W~B Mori.
\newblock One-to-one direct modeling of experiments and astrophysical
  scenarios: pushing the envelope on kinetic plasma simulations.
\newblock {\em Plasma Physics and Controlled Fusion}, 50(12):124034, 2008.

\bibitem{Ryutov:2012}
D~D Ryutov, N~L Kugland, H~S Park, C~Plechaty, B~A Remington, and J~S Ross.
\newblock Basic scalings for collisionless-shock experiments in a plasma
  without pre-imposed magnetic field.
\newblock {\em Plasma Physics and Controlled Fusion}, 54(10):105021, 2012.

\bibitem{0741-3335-52-12-124027}
C~K Li, F~H S{\'e}guin, J~A Frenje, {\em et al.}.
\newblock Diagnosing indirect-drive inertial-confinement-fusion implosions with
  charged particles.
\newblock {\em Plasma Physics and Controlled Fusion}, 52(12):124027, 2010.

\bibitem{Li:2009}
C.~K. Li, F.~H. S\'eguin, J.~A. Frenje, R.~D. Petrasso, P.~A. Amendt, R.~P.~J.
  Town, O.~L. Landen, J.~R. Rygg, R.~Betti, J.~P. Knauer, D.~D. Meyerhofer,
  J.~M. Soures, C.~A. Back, J.~D. Kilkenny, and A.~Nikroo.
\newblock Observations of electromagnetic fields and plasma flow in hohlraums
  with proton radiography.
\newblock {\em Phys. Rev. Lett.}, 102:205001, May 2009.

\bibitem{Kato:2010}
	Tsunehiko N. Kato and Hideaki Takabe.
	\newblock Electrostatic and electromagnetic instabilities associated with electrostatic shocks: Two-dimensional particle-in-cell simulation.
	\newblock {\em Phys. Plasmas}, 17:032114, 2010.

\bibitem{davidson:317}
Ronald~C. Davidson, David~A. Hammer, Irving Haber, and Carl~E. Wagner.
\newblock Nonlinear development of electromagnetic instabilities in anisotropic
  plasmas.
\newblock {\em Physics of Fluids}, 15(2):317--333, 1972.

\bibitem{berger:3}
R.~L. Berger, J.~R. Albritton, C.~J. Randall, E.~A. Williams, W.~L. Kruer,
  A.~B. Langdon, and C.~J. Hanna.
\newblock Stopping and thermalization of interpenetrating plasma streams.
\newblock {\em Physics of Fluids B: Plasma Physics}, 3(1):3--12, 1991.


\end{thebibliography}
% \phantom{x\\x\\x\\}

\end{document}